\documentclass[preprint,floats,aps,showpacs,groupaddress]{revtex4}
\usepackage{amssymb}
\usepackage{latexsym}
\usepackage[dvips]{graphicx}
\usepackage{amsmath}
\usepackage[mathcal]{eucal}
\usepackage{graphicx}
\usepackage{dcolumn}
\usepackage{amsmath}
\usepackage{amsfonts}
\usepackage{bm}

\usepackage{epsfig}

\newcommand{\be}{\begin{equation}}
\newcommand{\ee}{\end{equation}}
\newcommand{\bea}{\begin{eqnarray}}
\newcommand{\eea}{\end{eqnarray}}

\newcommand{\ri}{\mbox{i}}
\newcommand{\re}{\mbox{e}}

\begin{document}

\title{Analysis of the Quasiparticle Spectral Function in the Underdoped Cuprates}

\author{ M. Khodas }
\affiliation{ Department of  Condensed Matter Physics and Material Science, Brookhaven National Laboratory, Upton, NY 11973-5000, USA}
\author{H.-B. Yang}
\affiliation{ Department of  Condensed Matter Physics and Material Science, Brookhaven National Laboratory, Upton, NY 11973-5000, USA} 
\author{J. Rameau}
\affiliation{ Department of  Condensed Matter Physics and Material Science, Brookhaven National Laboratory, Upton, NY 11973-5000, USA} 
\author{P. D. Johnson}
\affiliation{ Department of  Condensed Matter Physics and Material Science, Brookhaven National Laboratory, Upton, NY 11973-5000, USA} 
\author{ A.M. Tsvelik}
\affiliation{ Department of  Condensed Matter Physics and Material Science, Brookhaven National Laboratory, Upton, NY 11973-5000, USA}
\author{  T. M. Rice  }
\affiliation{ Department of  Condensed Matter Physics and Material Science, Brookhaven National Laboratory, Upton, NY 11973-5000, USA}
\affiliation{ Institut f\"{u}r Theoretische Physik, ETH Z\"{u}rich,CH-8093 Z\"{u}rich, Switzerland }


\begin{abstract}

We applied the approach of K.-Y. Yang, T. M. Rice and F.-Ch. Zhang \cite{yrz} (YRZ)
 to analyze the high resolution angular resolved photo-emission spectroscopy (ARPES) data in BiSCO obtained recently at Brookhaven.
In the YRZ ansatz a constant RVB gap is assumed which leads to Luttinger zeros along the AFBZ and four Fermi pockets centered on the nodal directions. We relax the assumption of a constant RVB gap function, treating it as a Ising order parameter accompanied by thermal fluctuations. If these thermal fluctuations are very strong leading to strictly short range correlations in the spatial dependence of the RVB gap, then the reconstruction of the Fermi surface into pockets will not survive.
We examined the intermediate case of critical fluctuations leading to a power law falloff of the RVB gap correlations. To this end we followed the analysis recently developed by two of us to treat the effect of the power law correlations in the antiferromagnetic 2-dimensional xy model on the single electron GreenÕs function. 
   The partial truncation of the Fermi surface to form pockets survives in the presence of power law correlations. The linewidth of the quasiparticle peaks increases with the exponent of the power law correlations. If this exponent is set at the value of critical fluctuations in the 2-dimensional Ising model a relatively small linewidth is obtained. If this exponent is doubled, a linewidth comparable to the values found in recent ARPES experiments on underdoped BiSCO (Tc = 65 K)  at T = 140K was obtained. The anisotropic suppression of  the quasiparticle peak around the Fermi pockets hides the back side of the pockets closest to the AFBZ to give essentially Fermi arcs seen experimentally.

\end{abstract}

\maketitle

\sloppy
\section{Introduction}

 The underdoped regime of the cuprates remains the greatest mystery of condensed matter physics. Its resolution may have very profound influence on our understanding of strongly correlated phenomena. For several years after their discovery it was believed that the cuprates had a large hole like Fermi surface (FS) which area is proportional to $1-x$ with $x$ being concentration of the holes. This was celebrated as a triumph of the Luttinger theorem which apparently held even in such strongly correlated systems. Gradually more and more conflicting experimental evidence have emerged. Now it is overwhelmingly  obvious that in their normal state the underdoped cuprates have  a FS consisting of  either finite arcs or pockets centered around nodes of the Brillouin zone. The evidence for arcs comes from ARPES experiments \cite{arcs1},\cite{arcs}, and the evidence for pockets comes from both recent ARPES data \cite{pdj1} and most conclusively from quantum oscillation measurements in high magnetic fields \cite{oscillations},\cite{pockets1}.

 As far as the theory is concerned, there are two camps - traditional and radical ones. The traditional camp attempts to explain the phenomenon within the familiar
 paradigm of physics of weakly interacting systems. The radical camp argues that the doped Mott insulators such as cuprates display essentially nonperturbative features which do not occur in weakly interacting systems. The first question to ask is whether this phenomenon is a manifestation of some new physics characteristic for strongly correlated systems or can be understood within the traditional paradigm.

 In the traditional scenario  the pockets originate from the Fermi surface reconstruction  occurring in the presence of some density wave order parameter. This order parameter is likely to compete with the superconductivity and several candidates have been suggested. However, with the exception of LBCO at special doping no static ordering leading to increase of the unit cell has been  observed. This difficulty is explained away with a statement  that for FS reconstruction to occur it is sufficient to have a fluctuating order with a sufficiently large  correlation length. Detailed calculations in support of this scenario based on the so-called spin-fermion model can be found  in the paper by Chubukov and Morr \cite{chubmorr}  and in a more recent paper by Sedrakyan and Chubukov \cite{sedr}.  As the fluctuating order parameter the authors have taken  commensurate spin density wave (SDW). This choice leads to certain difficulties. 
 \begin{itemize}
 \item
 Perhaps, the most serious one comes from the fact that fluctuating SDW assumes predominance  of soft modes in the spin excitation spectrum. However, the pseudogap phenomenon is associated primarily with  the decrease of magnetic susceptibility which speaks against such a scenario.  What is important is that such a decrease  has been  observed in clean stoichiometric cuprates such as YBa$_2$Cu$_4$O$_8$ \cite{japan} and HgBa$_2$CuO$_{4 + \delta}$ \cite{bobroff},\cite{itoh}.  This excludes an interpretation of the spin gap as disorder phenomenon.
 Static CDW or SDW order would also lead to a splitting of NMR lines. Meanwhile, such effects have not been seen in the aforementioned   stoichiometric cuprates. To the contrary, the cited experiments show narrow NMR lines on different oxygen sites and a clear evidence of the spin gap in the temperature dependence of the Knight shift and the Gaussian spin-echo decay rate.  It is true that due to the technical problems related to the surface preparation ARPES and STS experiments on thee materials have not been done. This leaves a possibility to argue that there are no Fermi pockets there. This is, however, a narrow escape route since in all other respects these materials are not different from other cuprates.
 \item
 The other difficulty is related to the fact that the  choice of commensurate SDW places the centers of the pockets squarely at $(\pm \pi/2,\pm \pi/2)$ points of the Brillouin zone (Fig. 1). This   seems to be incompatible with the recent high resolution ARPES measurements \cite{pdj1}.

 \item
 The latter  difficulty does not go away even if one invokes incommensurate spin density wave suggested in \cite{moon},\cite{demler}:
 \bea
 && {\bf S} = \Re e[{\bf N}_1\re^{\ri{\bf Q}_1{\bf r}} + {\bf N}_2\re^{\ri{\bf Q}_2{\bf r}} ], \nonumber\\
 &&  {\bf Q}_1 = (\pi - \delta,\pi), ~~ {\bf Q}_2 = (\pi,\pi-\delta)
 \eea
 Although such SDW is naturally conjectured from the neutron experiments, which observe incommensurate low energy peaks in dynamical magnetic susceptibility \cite{tranquada}, in such choice would not do the job. Due to the Umklapp processes there are not four, but eight hot spots on the bare (large) FS connected by the wave vectors  $(\pm \pi \pm \delta,\pm \pi \pm \delta)$. This arrangement again produces the electron spectral weight symmetric with respect to the magnetic Brillouin zone boundary.

 \end{itemize}

\begin{figure}[h]
\includegraphics[width=0.5\columnwidth]{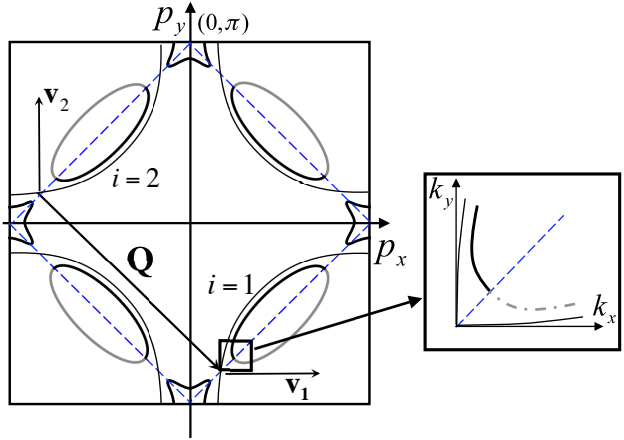}
\caption{Formation of a  Fermi pocket in the mean field Fermi surface reconstruction caused by the SDW ordering. The bare Fermi surface (thin solid line) is not nested.
The dashed line is the magnetic Brillouin zone boundary.
Two subbands with $i = 1,2$ are connected by the anti-ferromagnetic wave vector $\mathbf{Q} = (\pi,\pi )$.
 \label{fig:FS1} }
\end{figure}

 Nontraditional scenarios directly appeal to strongly correlated physics and suggest that in strongly correlated systems such as doped Mott insulators FS may be small incorporating only holes and this does not require any ordering. 
 Dzyaloshinskii reminded us that the Luttinger theorem, in fact, does not relate the particle density to the FS volume \cite{loshad} (see also \cite{chubdash}). 
 The latter relation exists only in weakly correlated systems where the electron self energy $\Sigma(\omega,{\bf k})$ does not have singularities at zero frequency. In strongly correlated systems where $\Sigma(\omega =0, {\bf k})$ goes to infinity at certain surface in momentum space, the volume inside of this surface contributes to the particle density. 
 A Mott insulator is an obvious example of an insulating  state with a half filled Brillouin zone which seemingly violates the Luttinger theorem (in its traditional, or rather, according to Dzyaloshinskii, confused understanding).

This idea was illustrated by Essler and Tsvelik \cite{EssTsv} who considered a strongly correlated model which allowed a controlled approximation. 
This was a model of Hubbard chains coupled by a long range tunneling (the inverse tunneling radius $\kappa$ served as a small parameter of the theory). Strong intrachain correlations were taken into account non-perturbatively and the interchain tunneling and exchange interaction processes were considered in a controlled fashion by RPA and perturbation theory in $\kappa$. The salient feature of this model was existence of a small FS in the form of hole and electrons pockets. 
These pockets existed in a finite temperature range; below a certain temperature the model undergoes a magnetic ordering. The main purpose of this  model was to illustrate  the principle.  
Namely the lack of the relation between the carrier density and the volume of the FS.
To apply these ideas to the cuprates one needs to study more realistic models.  This was done using Cluster Dynamical Mean Field (CDMF) approach;
states with small FS have been found (see, for instance, \cite{kotliar1},\cite{kotliar2},\cite{Imada}).

In principle, the radical scenario does not assume an extreme proximity to ordered states, though the model of Essler and Tsvelik has gapless collective excitations. Konik, Rice and Tsvelik (KRT) \cite{krt} obtained pockets in  a model of coupled ladders which does not have gapless modes though it is  fair to say that this model has two sites per unit cell and hence the translational invariance was already broken. Based on these results Yang, Rice and Zhang (YRZ) \cite{yrz} conjectured a phenomenological expression for the single particle Green's function which turned out to be highly successful in describing the ARPES and other data in underdoped cuprates, e. g. see \cite{RiceJohn},\cite{carbotte}:
\bea
G(\omega,{\bf k}) = \Big[(\omega +\ri 0) - \epsilon({\bf k}) - \frac{|\Delta({\bf k})|^2}{(\omega +\ri 0) - \xi_0({\bf k})}\Big]^{-1} , \label{yrz}
\eea
where $\Delta({\bf k}) = (\cos k_x - \cos k_y)\Delta$ and $\xi_0 = t_0(\cos k_x + \cos k_y)$ and $\epsilon({\bf k}) = t_1(\cos k_x + \cos k_y) + t_2\cos k_x\cos k_y - \mu$ describes the bare band. 
The function  $\Delta({\bf k})$ is the main input in the theory and requires explanation. 
In the KRT model the parent state is the so-called $d$-Mott state of two-leg ladder which has no local order parameter. Instead, at $T=0$ it has a topological order and $\Delta$ is the quasiparticle gap. It is not {\it apriori} clear what is the status of this function in two-dimensional case. One can imagine that it becomes a fluctuating order parameter and this is the option we explore in this paper.

 An expression similar to (\ref{yrz}) has recently been obtained by Qi and Sachdev \cite{subir} on the basis on the theory of algebraic charge liquid (ACL) suggested by Moon and Sachdev \cite{moon}. 
 The presence of an ACL places this theory into the radical camp. 
 The difficulty is that the  ACL state  has  gapless collective nonmagnetic excitations which presence would presumably be acutely felt in the thermodynamics. 
 As a matter of fact, the theory of the pocket formation suggested in \cite{subir} is not that different from \cite{EssTsv}. 
 In both cases the parent high energy state is fractionalized; the excitations are spinless particles with charge $e$ and neutral particles with spin S=1/2. 
 Their bound states are quasiparticles with a small FS. 
 An essential difference comes from the status of the parent state. 
 For Hubbard chain this state is well established through the exact solution of this model. 
 It is less certain with ACL. 
 The only example of such a state  has been found in  the model which physical parameters are dramatically different from those for the cuprates \cite{sandvik}. 
 Ironically, this is a situation typical for the cuprate physics: those models which allow rigorous treatment do not apply and those models which may apply do not allow rigorous treatment. 
 The other difference comes from the influence of gapless excitations. 
 In \cite{EssTsv} the interaction of quasiparticles with gapless collective modes strongly renormalizes the self energy resulting in a marginal Fermi liquid state. 
 In \cite{subir} the interaction between gapless gauge field excitations and quasiparticles is weak and the state is essentially a Fermi liquid.
 
 \section{Experimental Data and Fitting Procedure}

   In this paper we report an attempt to fit the ARPES data for BiSCCO using a modified form of YRZ spectral density. The ARPES spectra shown in this paper were recorded on beamline U13UB at the NSLS using a
Scienta SES2002 electron spectrometer. The samples are cleaved and measured in a UHV
chamber, where the base pressure is maintained at $5
\times10^{-11}$ Torr. Spectra are
recorded in the form of spectral intensity plots which represent a map of the intensity
as a function of binding energy and angle of emission or momentum. The energy and angular
resolution used in these studies was 15 meV and 0.1$^{\circ}$ respectively.   The experimental curves presented by dashed lines on Figs. \ref{fig:fit1}, \ref{fig:fit2},  \ref{fig:fit3}, \ref{fig:fit4},  have been obtained from the data by the deconvolution procedure described in \cite{pdj1}. This procedure removes a substantial part of the broadening caused by the experimental resolution effects. 
The spectral intensity may be sampled at a given binding energy as a function of momentum,
also known as momentum distribution curves (MDC).   
The Bi-2212 crystals were grown by the floating zone method. 
To get the under-doped BSCCO single crystals, the optimally
doped material was annealed under vacuum at 550 $^{\circ}$C for 48 hours. 
The $T_{c}$ of underdoped single crystals is 65 K, which was determined using SQUID magnetometry. The measurements were performed at fixed temperature T =140K.

 \begin{figure}[ht]
\epsfxsize=0.5\textwidth
\epsfbox{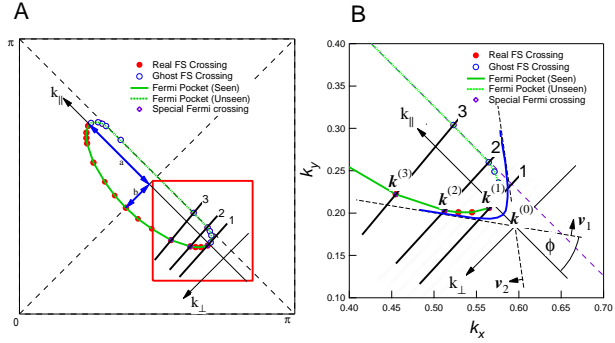}
\caption{ (a) The hole Pocket can be approximated as an ellipse with semiaxes $a$ and $b$.
The rotated system of coordinates $(k_{\perp},k_{\parallel})$, with $k_{\perp}$ measured along the nodal direction
is used to present the data.
(b) The zoom of the tip of the pocket region which contains the cuts 2 and 3. 
Points $\mathbf{k}^{(2)}$ and $\mathbf{k}^{(3)}$ on the pocket boundary are characterized by maximal ARPES intensity in MDC data along cut 2 and 3. 
Sufficiently close to the tip of the pocket, $\mathbf{k}^{(1)}$ it can be approximated by the hyperbolic curve asymptotically approaching two lines symmetric with respect to the $k_{\parallel}$ direction and crossing at the 
point $\mathbf{k}^{(0)}$.
The hyperbola touches the pocket and represent accurately the front side of the pocket.
The back side of the pocket gets shifted away from the line $k_{\perp}=0$.
This shift of the ghost FS is of little consequence as the ARPES intensity is very small at the back side of the pocket.   
The angle $\phi$ formed by the bare Fermi surface with the direction along $k_{\parallel}$ is determined by fitting the experimental data taken along cut 2 and cut 3. 
Vectors $\mathbf{v}_{1}$ and $\mathbf{v}_{2}$ are the velocities at the real and ghost FS respectively. 
 \label{fig:Cut} }
\end{figure}

The logic of our approach is as follows. 
Our goal is to check whether fluctuation effects are compatible with the formation of FS pockets. 
To this end we perform a kind of stress test on the YRZ and similar radical scenarios. 
In accordance with these scenarios we will follow the approach taken in \cite{khodas} and consider  the spectral function of quasiparticles moving in a slowly varying order parameter field. 
Further we assume that the fluctuations are as weak as  possible in the  absence of long range order. 
This means fluctuations with longest possible correlations length, i.e. critical ones. 
At finite temperatures critical fluctuations are also classical. 
So we assume that the system in question has a wide critical region where the order parameter fluctuations have power law correlations. 
In this way we consciously underestimate the effects of fluctuations. 
However,  if we manage to fit the data, this would mean that these radical approaches have a chance. 
If not a more radical scenario is needed. 
As far as the order parameter is concerned, we believe that the considerations described above exclude most of the explanations with a possible  exception of \cite{subir}.  
However, the latter theory also points to weak fluctuation effects.

To fit the experimental data we adopt the approach of Ref.~\cite{khodas} to treat classical fluctuations in YRZ theory.
In Ref.~\cite{khodas} the model of electrons in two subbands of opposite spins with dispersion relations 
$\epsilon_{1}(\mathbf{k}) =\epsilon(\mathbf{k})  $, and $\epsilon_{2}(\mathbf{k}) =\epsilon(\mathbf{k}+\mathbf{Q})  $
coupled by the antiferromagnetic order parameter has been studied, see Fig.~\ref{fig:FS1}. 
The effective Hamiltonian in the two subband space describing this situation can be written as
\begin{align}								\label{hmf}
H = \epsilon_{1}(\mathbf{k})\frac{ 1+ \sigma_{z}}{2} +  \epsilon_{2}(\mathbf{k})\frac{ 1- \sigma_{z}}{2}
+ \Delta(\mathbf{k})\frac{\sigma_{+}}{ 2 } +  \Delta^{*}(\mathbf{k})\frac{\sigma_{-}}{2}\, ,
\end{align}
where $\sigma_{i}$ are Pauli matrices acting in the two subbands space, and
$\sigma_{\pm} = \sigma_{x} \pm \ri \sigma_{y}$.
This Hamiltonian with constant order parameter $\Delta(\mathbf{k})$ 
corresponds to the mean field solution. 
In this limit of no fluctuations the Fermi surface is reconstructed and the Green function reads
\bea
G_{mf}(\omega,{\bf k}) = \Big[(\omega +\ri 0) - \epsilon_{1}({\bf k}) - \frac{|\Delta({\bf k})|^2}{(\omega +\ri 0) - \epsilon_2({\bf k})}\Big]^{-1} \, . \label{mf}
\eea
To study the effects of fluctuations, it has been assumed in \cite{khodas} that the spin magnetization either 
i) has a strong easy plane anisotropy and lies in the $XY$-plane or
ii) has a strong easy axis anisotropy and is directed along the z axis.
Most importantly the regime of critical fluctuation has been studied in both cases.
In the first case the fluctuations are critical in the whole range of temperatures below a BKT transition.
While in the second case they are critical only at the transition.
The self-energy calculated to second order in perturbation theory in $\Delta$ yields for the Green function
\begin{align}							\label{SDWsecond}
G_{2}^{-1}  = \omega - \epsilon_{1}( \mathbf{k} ) + \frac{ \ri \Delta(\mathbf{k})^{2} \ell^{2d} \Gamma( 1 - 2d )}{ (-\ri (\omega - \epsilon_{2}(\mathbf{k}) ))^{1 -2d} }\, ,
\end{align}
where $d$ is the scaling dimension of $\Delta$, and $\ell$ is the ultraviolet scale of the order of the lattice spacing.
The new energy scale entering the Green function in Eq.~\eqref{SDWsecond} is 
$
T_{K} = \Delta (\Delta \ell)^{d/(1-d)}
$.
The self energy in Eq.~\eqref{SDWsecond} is regular at the mass shell, $\omega = \epsilon_{1}(\mathbf{k})$.
However at higher orders it contains singular terms at both the mass shell, $\omega = \epsilon_{1}(\mathbf{k})$ and at the ghost mass shell   $\omega = \epsilon_{2}(\mathbf{k})$. 
These singularities were systematically resummed in Ref.~\cite{khodas}.
We focus the discussion on the result obtained close to the mass shell, where most of the spectral weight accumulates, and comment on the behavior at the ghost Fermi surface later.
In the former case the result reads
\begin{align}              \label{SDWshell1}
G^{-1} =  G_{2}^{-1}  + \frac{ 2d \ell^{4d} \Gamma^{2}(2-2d) \Delta^{4}  }{ (- \ri (\omega - \epsilon_{2}(\mathbf{k})  )  )^{4 - 4d}   }G_{2}^{-1} 
\ln\left( \frac{ G_{2}^{-1} }{ \omega - \epsilon_{2}(\mathbf{k}) } \right)
 + \frac{ 2d \ri \ell^{6d} \Gamma^{2}(2-2d) \Gamma(1 - 2d)  \Delta^{6}  }{ (-\ri (\omega - \epsilon_{2}(\mathbf{k}) ) )^{5 - 6d} }\, ,
\end{align}
where $G_{2}$ is given by Eq.~\eqref{SDWsecond}.
Equation \eqref{SDWshell1} is valid in the limit $|\omega - \epsilon_{1}(\mathbf{k})|  \lesssim  |\omega - \epsilon_{2}(\mathbf{k})| $,
and $ T_{k} \ll |\omega - \epsilon_{2}(\mathbf{k})| $.

%

As we have said, the expression \eqref{SDWshell1} is quite general. 
If the bare Fermi surface is considered as the first subband , $\epsilon_{1}(\mathbf{k}) = \epsilon(\mathbf{k})$,
and we take the magnetic zone boundary for the second one,  $\epsilon_{2}(\mathbf{k}) = \xi_{0}(\mathbf{k})$
Eq.~\eqref{yrz} is obtained in the mean field, Eq.~\eqref{mf}.
With this correspondence, Eqs.~\eqref{SDWsecond}, \eqref{SDWshell1} become
\begin{align}							\label{second}
G_{2}^{-1}  = \omega - \epsilon( \mathbf{k} ) + \frac{ \ri \Delta(\mathbf{k})^{2} \ell^{2d} \Gamma( 1 - 2d )}{ (-\ri (\omega - \xi_{0}(\mathbf{k}) ))^{1 -2d} }\, ,
\end{align}
\begin{align}              \label{shell1}
G^{-1} =  G_{2}^{-1}  + \frac{ 2d \ell^{4d} \Gamma^{2}(2-2d) \Delta^{4}  }{ (- \ri (\omega - \xi_{0}(\mathbf{k})  )  )^{4 - 4d}   }G_{2}^{-1} 
\ln\left( \frac{ G_{2}^{-1} }{ \omega - \xi_{0}(\mathbf{k}) } \right)
 + \frac{ 2d \ri \ell^{6d} \Gamma^{2}(2-2d) \Gamma(1 - 2d)  \Delta^{6}  }{ (-\ri (\omega - \xi_{0}(\mathbf{k}) ) )^{5 - 6d} }\, .
\end{align}
These equations have a meaning of the phenomenological   YRZ function   modified by fluctuations. 
In this modification we treat $\Delta$ in (\ref{yrz}) as an Ising-like order parameter and assume that its fluctuations are classical and critical. 
The choice of Ising order parameter dictates the value of the scaling dimension of $\Delta$ as $d=1/8$. 
However in the absence of full microscopic derivation of Eq.~\eqref{yrz} the scaling dimension $d$ should be really considered as a fitting parameter with fluctuations enhanced for larger vales of $d$.
In fact, we have found that $d=1/4$ more accurately describes our data. 
We additionally assume that fluctuations remain critical in a wide temperature range as is the case for the $XY$ model below the BKT transition.
This assumption probably introduces a bias towards order thus making the spectral lines narrower than they must be. 
We remind the reader that the scenario described in \cite{subir}, where quasiparticles have $\sim \omega^2 + T^2$ decay rate, is likely to lead to even sharper peaks. 

Below we introduce simplifying approximations to the dispersion to facilitate the comparison of Eqs.~\eqref{second}, \eqref{shell1} with the experimental data. 
Close to the tip of the pocket the bare dispersion $\epsilon(\mathbf{k})$ can be linearized,
$\epsilon(\mathbf{k}) \approx \mathbf{v}_{1} (\mathbf{k} -\mathbf{k}^{(0)})$.
Here the point $\mathbf{k}^{(0)}$ is the crossing of the bare Fermi surface $\epsilon(\mathbf{k})=0$ 
with the line passing through the two tips of the pocket,  see 
Fig.~\ref{fig:Cut}. 
In what follows it is convenient to choose $\mathbf{k}^{(0)}$ as the origin of
coordinate frame in the Brillouin zone with axes $k_{\parallel}$ and $k_{\perp}$ directed parallel and perpendicular to the magnetic zone boundary respectively,  see 
Fig.~\ref{fig:Cut}.
In the above coordinate frame the velocity $\mathbf{v}_{1} =  ( v \sin \phi, -v \cos \phi )$ is specified by the angle $\phi$ it forms with the diagonal direction,  see Fig.~\ref{fig:Cut}.
For simplicity we also modify the line of zeros of the Green function in Eq.~\eqref{yrz},
$\xi_{0}(\mathbf{k}) \rightarrow  \mathbf{v}_{2} \mathbf{k} $ with $\mathbf{v}_{2} =  ( v \sin \phi, v \cos \phi )$.
We stress that although this modification leads to a deformation of  the back side of the pocket which is discussed later,
it does not change appreciably the behavior of the Green function at the mass shell which is the focus of current discussion. 
In the linear approximation the mean field, pocket is approximated by the hyperbolic curve 
asymptotically approaching lines perpendicular to velocities $\mathbf{v}_{1,2}$,  see 
Fig.~\ref{fig:Cut}.
The requirement that the hyperbola touches the pocket and has the same curvature at the tip imposes a relation between the distance $|\mathbf{k}^{(1)} - \mathbf{k}^{(0)}|$ and the angle $\phi$. 
The latter is the only free parameter  for fixed $d$ which has to be determined by the fitting procedure described in App.~\ref{app:fit}.

We express the result~\eqref{shell1} in the universal form
$ G = T_{K}^{-1}\bar{G} ( \bar{\mathbf{k}}, \bar{\omega})$ through 
dimensionless momentum,
$\bar{\mathbf{k}} = v \mathbf{k} / T_{K}$ and energy $\bar{\omega} = \omega / T_{K}$, with
\begin{align}              \label{shell1bar}
\bar{G}^{-1} = & \bar{G}_{2}^{-1}  +
\frac{ 2d \Gamma^{2}(2-2d)  }{ (- \ri (\bar{\omega} - \bar{k}_{\perp}\cos \phi -\bar{k}_{\parallel}\sin \phi ))^{4 - 4d} }\bar{G}_{2}^{-1}
\ln\left( \frac{ \bar{G}_{2}^{-1} }{ \bar{\omega} - \bar{k}_{\perp}\cos \phi -\bar{k}_{\parallel}\sin \phi } \right)
\notag \\
& + \frac{ 2d \ri \Gamma^{2}(2-2d) \Gamma(1 - 2d)   }{(-\ri (\bar{\omega} - \bar{k}_{\perp}\cos \phi -\bar{k}_{\parallel}
\sin \phi))^{5 - 6d} }\, ,
\end{align}
where 
%
%
\begin{align}
\bar{G}_{2}^{-1}  =
\bar{\omega} + \bar{k}_{\perp}\cos \phi -\bar{k}_{\parallel}\sin \phi + \frac{ \ri \Gamma( 1 - 2d )}{ (-\ri (\bar{\omega} - \bar{k}_{\perp}\cos \phi -\bar{k}_{\parallel}\sin \phi))^{1 -2d} }\, .
\end{align}

To compare our results with experiment we plot the spectral density, $A(\omega,k_{\parallel},k_{\perp})= - 1/\pi  \mathrm{Im} G(\omega,k_{\parallel},k_{\perp})$ 
for fixed $k_{\parallel}$ as a function of $k_{\perp}$ corresponding to MDCs taken along the cut across the pocket.
We first consider the FS, $\omega=0$. 
In Fig.~\ref{fig:fit1} we show the fit to the experimental data shown by dashed lines with the theory for 
the scaling dimension $d=0.125$. 
This value would correspond to a transition region of two-dimensional Ising model, or to the BKT transition in the $XY$ model.
The theoretical expression \eqref{shell1} used in the comparison describes the spectral density at the real FS. 
In Fig.~\ref{fig:fit1} it is plotted in the interval $k_{\perp} > 0$ which includes the real FS, or more generally the mass shell, $\omega = \epsilon(\mathbf{k})$.
Both the data and the fit are presented for cut 2 and cut 3 taken parallel to the nodal ($\Gamma$-X) direction, see Fig.~\ref{fig:Cut}  with two different values of $k_{\parallel}$. 
Positions of both peaks are fitted simultaneously. 
The angle $\phi$ and the ratio $k_{\perp}/ \bar{k}_{\perp}= T_{K}/v $ used to plot Figs.~\ref{fig:fit1},~\ref{fig:fit2},~\ref{fig:fit3},~\ref{fig:fit4} are obtained through the 
approximate procedure described in App.~\ref{app:fit}.
This gives an energy scale $T_{K}=54$meV. 
We notice that the obtained energy scale is of the order of pseudogap,  $T_{K} \approx T^{*}$ as expected.
Although the position of the peaks is captured by the fit  their width, is smaller than in the actual data. 
The main point is that one can clearly see from Fig.~\ref{fig:fit1} that the FS survives critical fluctuations.
For the negative energy $\omega=-45$meV corresponding to the hole states well below the FS the data is shown in Fig.~\ref{fig:fit2} (dashed line). 
The theoretical fits are plotted according to the same Eq.~\eqref{shell1} with previously obtained scale $T_{K}$.
Here again the peak position is reproduced quite closely, but theory yields peaks which are too narrow.
We argue that other sources of broadening such as Landau damping mentioned earlier might be responsible for the extra width of the signal.
To see if the FS survives in the presence of stronger fluctuations we consider larger values of the scaling dimension $d$. 
In the absence of detailed microscopic derivation of Eq.~\eqref{yrz} this parameter is considered as free. 
In Fig.~\ref{fig:fit3} we show the fit to the experimental data taken at the FS, $\omega = 0$
with $d=0.25$.
In this case the agreement is improved as the higher values of $d$ correspond to a stronger fluctuations, and in result give a broader MDC.
The same fitting procedure (see App.~\ref{app:fit}) yields a slightly different value for the energy scale,
$T_{K} \approx 67$meV.
In Fig.~\ref{fig:fit4} we show the fit with identical parameters but for a negative energy,
$\omega=-45$meV.
At this energy the fit is much less accurate, but it still reflects the basic observation that the FS withstand the critical fluctuations. 
To summarize, critical fluctuations slightly modify the shape of the pockets, introduce a finite smearing as is  reflected in the shape of MDCs. 
But most importantly their effect is not strong enough as to restore the FS topology, and pockets survive.
Until now we have focused on the region $k_{\perp} > 0$ which contains the FS. 
In the present paragraph we discuss briefly the fate of the back side of the pockets lying in the complementary interval, $k_{\perp} < 0$.
The critical fluctuations tend to suppress the ghost FS, so that the spectral weight mostly accumulates  
at the front side of the pocket or close to the real FS, leading to Fermi arcs.  
We still discuss the Green function close to the ghost Fermi surface, 
$\omega \approx \xi_{0}(\mathbf{k})$ here for the sake of completeness.
Equation \eqref{shell1}  is not valid anymore in the interval $k_{\perp} < 0$ and has to be replaced.
In the Ref.~\cite{khodas} close to the ghost FS, $\omega \approx \epsilon_{2}(\mathbf{k})$, we have obtained the following result  
\begin{align}\label{SDWshell2}
G =  &
\frac{ 1 }{ \omega - \epsilon_{1}( \mathbf{k} ) } +
\frac{ \Delta^{2}( \mathbf{k} )\ell^{2d} \re^{\ri \pi d} \Gamma( 1\! -\! 2d )  }{ ( \omega - \epsilon_{1}( \mathbf{k} ))^{2 }  }  
 \left[ \omega\! -\! \epsilon_{2}(\mathbf{k})\! +\! \frac{ \ri \Delta^{2}( \mathbf{k} )   \ell^{2d}  \Gamma (1 -2d) }{(-\ri ( \omega - \epsilon_{1}( \mathbf{k} ) ))^{1 - 2d }  }  \right]^{\! - 1\! +\! 2d}.
\end{align}
The result in Eq.~\eqref{shell2} was obtained in the following way.
In the second order of perturbation theory in $\Delta$, Eq.~\eqref{SDWsecond}, the Green function acquires a correction singular at the ghost FS, $\omega = \epsilon_{2}(\mathbf{k})$. 
Higher orders corrections are even more singular.
Equation \eqref{SDWshell2} is result of careful resummation of the most singular terms  at $\omega = \epsilon_{2}(\mathbf{k})$.
It is therefore applicable in the limit $|\omega - \epsilon_{2}(\mathbf{k}) | \lesssim |\omega -\epsilon_{1}(\mathbf{k}) | $, and $ T_{k}  \ll  |\omega - \epsilon_{1}(\mathbf{k})| $.  
In exact similarity to the previously considered case of the FS in order to adopt the result \eqref{SDWshell2} to our present purposes one has to make a substitution $\epsilon_{1}(\mathbf{k}) = \epsilon(\mathbf{k})$, and
$\epsilon_{2}(\mathbf{k}) = \xi_{0}(\mathbf{k})$.
The resulting expression reads
\begin{align}\label{shell2}
G =  &
\frac{ 1 }{ \omega - \epsilon( \mathbf{k} ) } +
\frac{ \Delta^{2}( \mathbf{k} )\ell^{2d} \re^{\ri \pi d} \Gamma( 1\! -\! 2d )  }{ ( \omega - \epsilon( \mathbf{k} ))^{2 }  }  
 \left[ \omega\! -\! \xi_{0}(\mathbf{k})\! +\! \frac{ \ri \Delta^{2}( \mathbf{k} )   \ell^{2d}  \Gamma (1 -2d) }{(-\ri ( \omega - \epsilon( \mathbf{k} ) ))^{1 - 2d }  }  \right]^{\! - 1\! +\! 2d}.
\end{align}
In the mean field limit, $d=0$, the expression \eqref{shell2} goes over to the Eq.~\eqref{yrz} as expected.
At finite $d$ it expresses the effect of critical fluctuations on the YRZ ansatz at the ghost FS in much the same way as Eq.~\eqref{shell1} describes the modification of Eq.~\eqref{yrz} by critical fluctuations at the real FS.
The applicability conditions of Eq.~\eqref{shell2} break down sufficiently close to the tip of the pocket,
$\mathbf{k} \approx \mathbf{k}^{(1)}$ see Fig.~\ref{fig:Cut},  
and our results are expected to be less reliable in this region.
Indeed, the better theoretical fits are obtained for the cut 3 taken further away from the tip of the pocket than for the cut 2 in Figs.~\ref{fig:fit1}, \ref{fig:fit2}, \ref{fig:fit3} and \ref{fig:fit4}.

In order to enable the comparison of Eq.~\eqref{shell2} with experiment we repeat the steps leading to Eq.~\eqref{shell1bar}. 
Namely we rewrite Eq.~\eqref{shell2} in the form
$ G = T_{K}^{-1}\bar{G} ( \bar{\mathbf{k}}, \bar{\omega})$ with 
dimensionless momentum, $\bar{\mathbf{k}} = v \mathbf{k} / T_{K}$ and energy $\bar{\omega} = \omega / T_{K}$, 
\begin{align}\label{shell2bar}
\bar{G} =  & \frac{ 1 }{  \bar{\omega} + \bar{k}_{\perp} \cos \phi - \bar{k}_{\parallel} \sin \phi  } +
\frac{  \re^{\ri \pi d} \Gamma( 1\! -\! 2d )  }{ ( \bar{\omega} + \bar{k}_{\perp} \cos \phi - \bar{k}_{\parallel} \sin \phi   )^{2 }  }
\notag \\
& \times \left[ \bar{\omega} -\bar{k}_{\perp} \cos \phi - \bar{k}_{\parallel} \sin \phi  + \frac{ \ri \Gamma (1 -2d) }{(-\ri ( \bar{\omega} + \bar{k}_{\perp} \cos \phi - \bar{k}_{\parallel} \sin \phi ) )^{1 - 2d }  }  \right]^{ - 1 + 2d}.
\end{align}
The energy scale $T_{K}$ the angle $\phi$ in Eq.~\eqref{shell2bar} are identical to those in Eqs.~\eqref{shell1}, \eqref{shell1bar}.
We emphasize that in general the spectral density at the ghost Fermi surface tends
to be suppressed by fluctuations with most of the spectral weight found in the region $k_{\perp} > 0$, close 
to the Fermi surface.
We use the form of Eq.~\eqref{shell2bar} to obtain the spectra for $k_{\perp}<0$ and the results are included in Figs.~\ref{fig:fit1}, \ref{fig:fit2}, \ref{fig:fit3} and \ref{fig:fit4}.
The spectra for the two regions $k_{\perp}>0$ and $k_{\perp}<0$ mostly match each other.
The discontinuity at $k_{\perp}=0$ is smaller for cut 3 than for cut 2, see Fig.~\ref{fig:Cut}.
This is to be expected since as discussed above both Eq.~\eqref{shell1} used for 
$k_{\perp}>0$ and Eq.~\eqref{shell2} used for 
$k_{\perp}<0$ are better satisfied away form the tip of the pocket.


 \begin{figure}[ht]
\epsfxsize=0.6\textwidth
\epsfbox{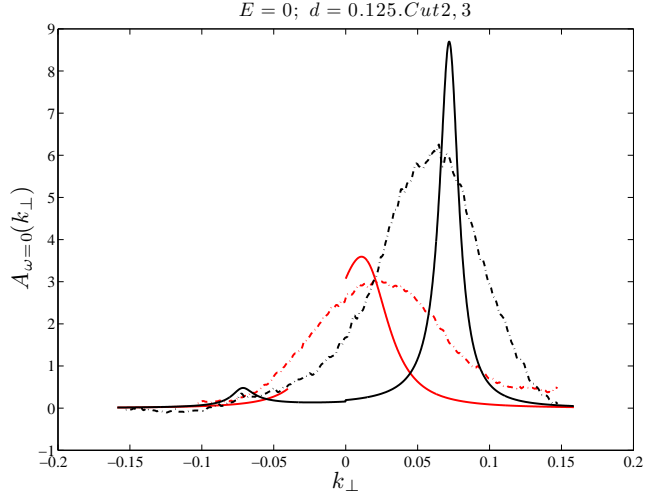}
\caption{Theoretical fits shown in solid line, to experimental data shown in dashed line taken along the Cut 2 (red) and Cut 3 (black) in Fig.~\ref{fig:Cut} at zero energy with $d=0.125$. 
The expression Eq.~\eqref{shell1bar} (Eq.~\eqref{shell2bar})  is used to plot the theoretical curves for $k_{\perp}>0$ ($k_{\perp}<0$).  
The fit in a subinterval of negative $k_{\perp}$ is not shown where the result Eq~\eqref{shell2bar} is not applicable.
 \label{fig:fit1} }
\end{figure}

 \begin{figure}[ht]
\epsfxsize=0.6\textwidth
\epsfbox{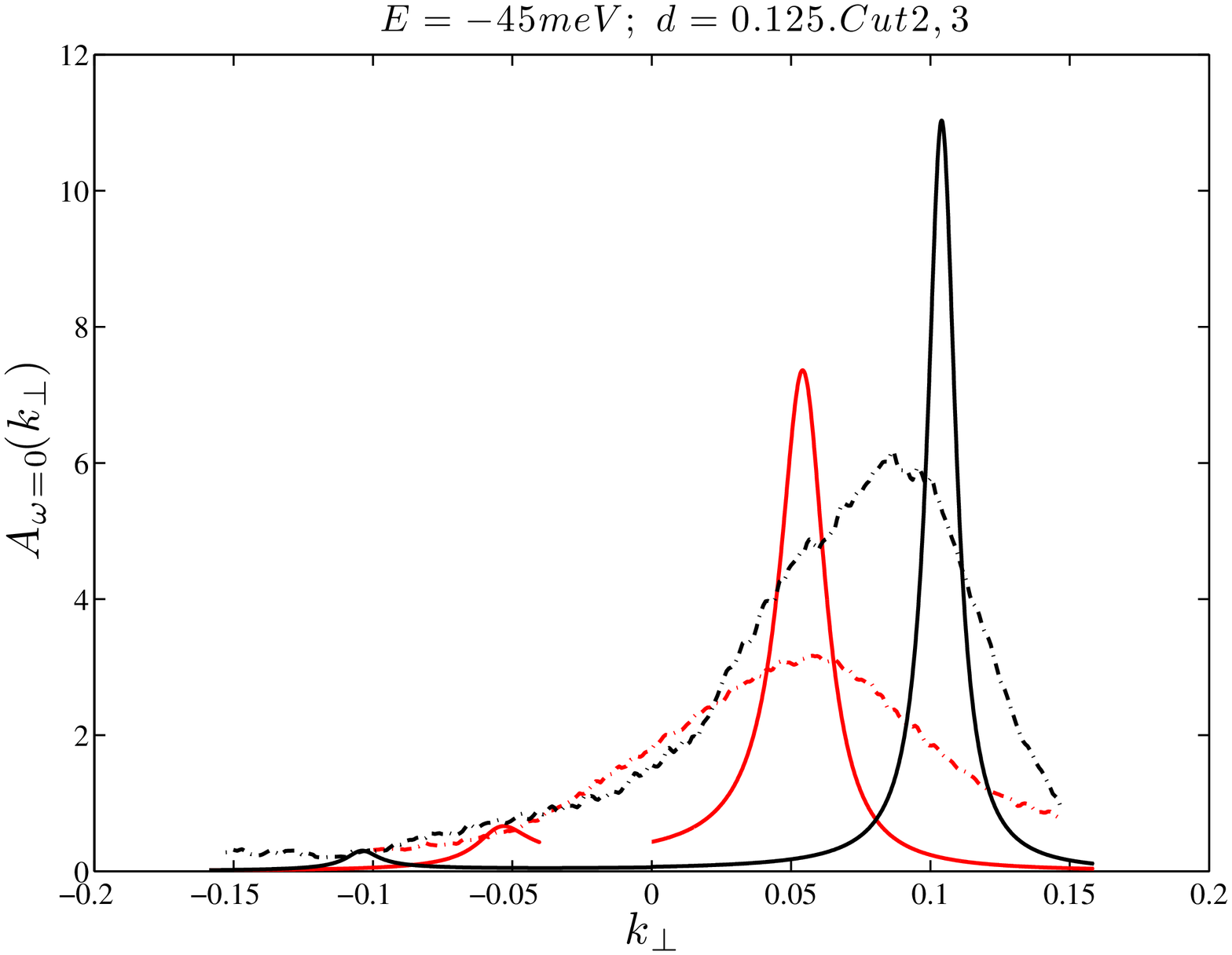}
\caption{Theoretical fits shown in solid line, to experimental data shown in dashed line taken along the Cut 2 (red) and Cut 3 (black) in Fig.~\ref{fig:Cut} at the energy $E = -45$meV with $d=0.125$. 
The expression Eq.~\eqref{shell1bar} (Eq.~\eqref{shell2bar})  is used to plot the theoretical curves for $k_{\perp}>0$ ($k_{\perp}<0$).  
The fit in a subinterval of negative $k_{\perp}$ is not shown where the result Eq~\eqref{shell2bar} is not applicable.
 \label{fig:fit2} }
\end{figure}

 \begin{figure}[ht]
\epsfxsize=0.6\textwidth
\epsfbox{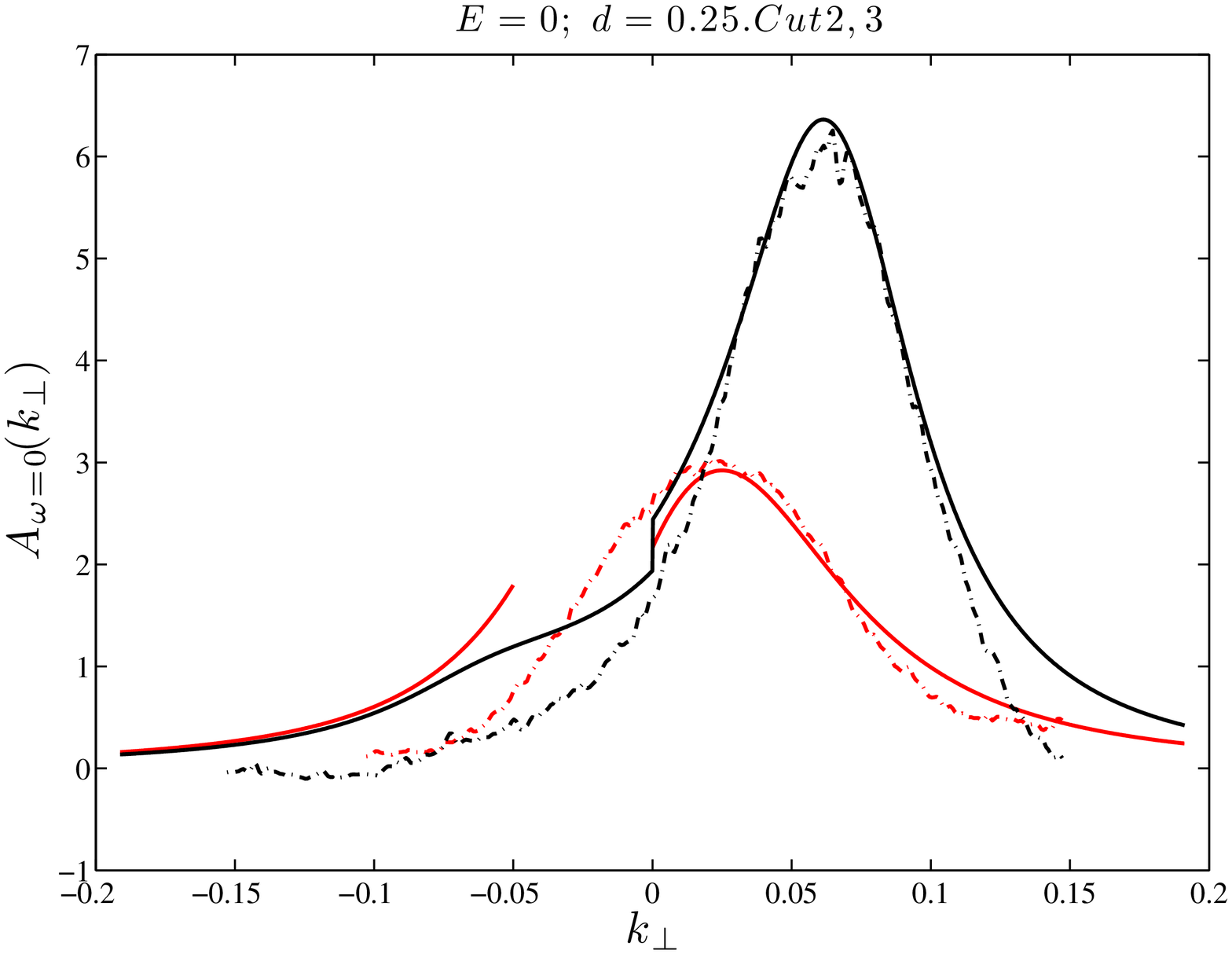}
\caption{Theoretical fits shown in solid line, to experimental data shown in dashed line taken along the Cut 2 (red) and Cut 3 (black) in Fig.~\ref{fig:Cut} at zero energy with $d=0.25$. 
The expression Eq.~\eqref{shell1bar} (Eq.~\eqref{shell2bar})  is used to plot the theoretical curves for $k_{\perp}>0$ ($k_{\perp}<0$).  
The fit in a subinterval of negative $k_{\perp}$ is not shown where the result Eq~\eqref{shell2bar} is not applicable.
 \label{fig:fit3} }
\end{figure}

 \begin{figure}[ht]
\epsfxsize=0.6\textwidth
\epsfbox{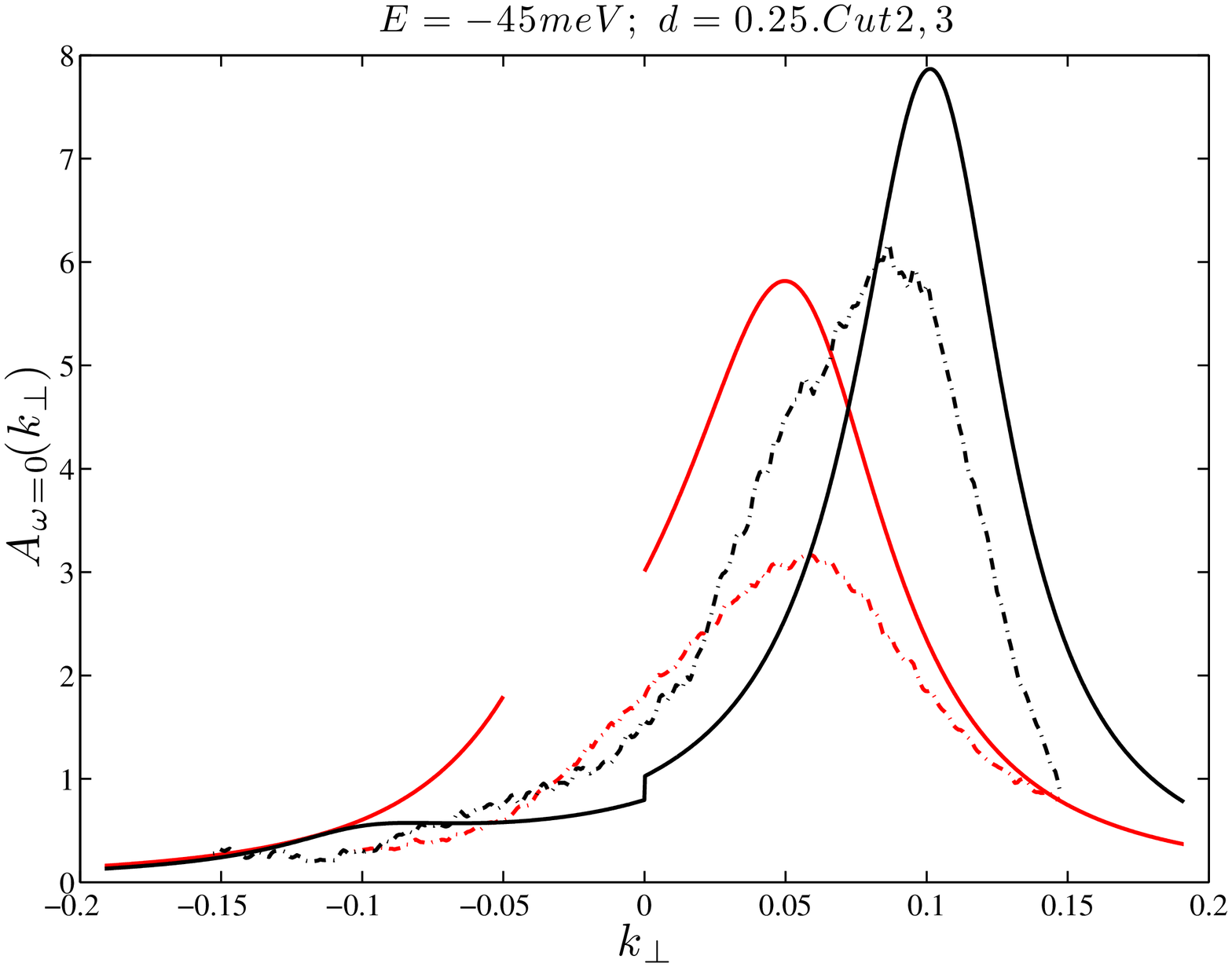}
\caption{Theoretical fits shown in solid line, to experimental data shown in dashed line taken along the Cut 2 (red) and Cut 3 (black) in Fig.~\ref{fig:Cut} at the energy $E = -45$meV with $d=0.25$. 
The expression Eq.~\eqref{shell1} (Eq.~\eqref{shell2})  is used to plot the theoretical curves for $k_{\perp}>0$ ($k_{\perp}<0$).  
The fit in a subinterval of negative $k_{\perp}$ is not shown where the result Eq~\eqref{shell2bar} is not applicable.
 \label{fig:fit4} }
\end{figure}

\section{Conclusions}

   In this paper we have examined the robustness of radical theories which have been proposed to describe  the anomalous properties of the  pseudogap state. We label these as radical because they are not derived from a broken symmetry order. In the YRZ case a phenomenological form for the single particle GreenÕs function is put forward in analogy to  the form that was derived for a two dimensional array of 2-leg Hubbard ladders near half-filling. In the ACL case \cite{subir} the key ingredient is the choice of the quantum disordering of a antiferromagnetic state in the presence of doped holes. In both cases the Fermi surface is partially truncated and a qualitative description of key experiments of the pseudogap state is obtained. An important open issue is the effect of thermal fluctuations and the robustness of the key properties in their presence.
     In the YRZ ansatz a constant RVB gap is assumed which leads to Luttinger zeros along the AFBZ and four Fermi pockets centered on the nodal directions. We relax the assumption of a constant RVB gap function, treating it as a Ising order parameter accompanied by thermal fluctuations. If these thermal fluctuations are very strong leading to strictly short range correlations in the spatial dependence of the RVB gap, then the reconstruction of the Fermi surface into pockets will not survive.
In this paper we examined the intermediate case of critical fluctuations leading to a power law falloff of the RVB gap correlations. To this end we followed the analysis recently developed by two of us to treat the effect of the power law correlations in the antiferromagnetic 2-dimensional xy model on the single electron electron GreenÕs function. 
   The main conclusion is that the partial truncation of the Fermi surface to form pockets survives in the presence of power law correlations. The linewidth of the quasiparticle peaks increases with the exponent of the power law correlations. If this exponent is set at the value of critical fluctuations in the 2-dimensional Ising model a relatively small linewidth is obtained. If this exponent is doubled, a linewidth comparable to the values found in recent ARPES experiments on underdoped BiSCO (Tc = 65 K)  at T = 140K was obtained. The anisotropic suppression of  the quasiparticle peak around the Fermi pockets hides the back side of the pockets closest to the AFBZ to give essentially Fermi arcs seen experimentally.
   We conclude that the main features of the Fermi surface reconstruction in a radical theory such as the YRZ ansatz, are robust against power law thermal fluctuations. 

We finally comment on the situation when there is no true order and the opposite case of an ordered state.
In the absence of true order the fluctuations are characterized by the finite correlation length.
The pockets  don't survive and the FS remains large and not truncated.
In the case of critical fluctuations studied in the present paper the self energy includes both real and imaginary part.
The real part modifies the dispersion of the quasi-particles leading to a FS reconstruction.
The imaginary part produces finite width of the spectral function.
The imaginary part is smaller than the real part by a factor of $d$.
In result, for $d < 1$ the FS pockets are well defined.
The situation is different in the absence of order: while the fluctuations still produce a finite broadening,
there is no real part leading to a FS truncation.
For that reason the pockets are destroyed in the absence of order.

The influence of thermal fluctuations on the spectral density in the ordered state has been studied recently in Ref.~\cite{sedr}.
The authors have employed the eikonal approximation and summed up the most singular corrections in the parameter $T/J |\log \epsilon |$, where $J$ is the exchange interaction in the Heisenberg model, and $\epsilon$ is the anisotropy parameter of a quasi-2D system. In the presence of long-range order the pockets have been found. In addition, some signatures of pockets remain even without long-range order. The fluctuations are strong enough in this case to redistribute the spectral weight evenly  between the real and the ghost FS. Our situation is intermediate in the sense that we have only quasi long range order, but the fluctuations have infinite correlation length. Our conclusion is that this situation is rather similar to the ordered case, i. e. the FS still has small pockets. The results of the Ref.~\cite{sedr} indicate that we have considered the strongest fluctuations possible consistent with small FS topology.

\acknowledgements
We are grateful to  A. V. Chubukov, R. M. Konik, I. A. Zaliznyak for valuable discussions.
We acknowledge support of the Center for Emergent Superconductivity,
an Energy Frontier Research Center supported by the US DOE,
Office of Basic Energy Sciences.
M. K. acknowledges support from BNL LDRD under Grant No. 08-002.

\begin{appendix}
\section{Fitting procedure}
\label{app:fit}
In this appendix we describe the fitting procedure used to estimate the characteristic energy scale $T_{K} $.
In our scheme the only fitting parameter is the  critical dimension $d$. 
For a given $d$ we approximate the Fermi surface by the hyperbolic curve touching the pocket at its tip, $\mathbf{k}=\mathbf{k}^{(1)}$, see Fig.~\ref{fig:Cut}.
obtained for the linearized bare dispersion. 
The bare Fermi surface is characterized by the angle $\phi$ it forms with the magnetic Brillouin zone boundary. 
The pocket in turn is approximated by an ellipse with major and minor semiaxes $a$ and $b$ respectively.

In order to fix $\phi$ we find  $\bar{k} _{\perp}^{(2)}$, and $\bar{k} _{\perp}^{(3)}$ by solving the equation $\mathrm{Re} G_{2}^{-1}(\omega=0,\bar{k} _{\perp}^{(i)},\bar{k} _{\parallel}^{(i)} )= 0$, $i=2,3$ numerically for different values of $\phi$. 
From the Fig.~\ref{fig:Cut} we deduce 
$ k^{(2)}_{\parallel} = k^{(1)}_{\parallel} + 0.11 a $, and
$ k^{(3)}_{\parallel} = k^{(1)}_{\parallel} + 0.36 a $.
Using the geometrical relation,
$1 = k_{\parallel}^{(1)}(a/b^{2}) \tan^{2}\phi$ 
and placing the tip of the hyperbolic curve at
$\bar{k}_{\parallel}^{(1)} \approx 1/ \sin \phi$ obtained form  
$ \mathrm{Re} G_{2}^{-1}(\omega=0,\bar{k} _{\perp}=0,\bar{k} _{\parallel} )= 0$ at the mean field $d=0$,
the ratio 
$\bar{k}_{\perp}^{(3)}/\bar{k}_{\perp}^{(2)} $ can be found.
The latter becomes equal to the experimental value 
$k_{\perp}^{(3)}/k_{\perp}^{(2)} $ at $\phi  \approx 0.21 \pi$ for $d = 0.125$.
In view of various approximations made in our procedure our estimate for $\phi$ requires a fine tuning
and the value $\phi \approx 0.23 \pi $ for $d=0.125$, and 
 $\phi \approx 0.22 \pi $ for $d=0.25$
better fit the data.  
The ratio  $k_{\perp}^{(2)} / \bar{k}_{\perp}^{(2)} = k_{\perp}^{(3)} / \bar{k}_{\perp}^{(3)} =T_{K}/v $
gives $T_{K} = 54$ meV  for $d = 0.125$ and $T_{K} = 67$ meV  for  and $d = 0.25$ respectively.
The scale $T_{K}$ is of the order of magnitude of the pseudogap energy $T^{*}$.

\end{appendix}


\begin{thebibliography}{99}
\bibitem{yrz} K.-Y. Yang, T. M. Rice and F.-Ch. Zhang, Phys. Rev. B{\bf 73}, 174501 (2006).
\bibitem{arcs1} A. Kanigel, M. R. Norman, M. Randeria, U. Chatterjee, S. Suoma, A. Kaminski, H. M. Fretwell, S. Rosenkranz, M. Shi, T. Sato, T. Takahashi, Z. Z. Li, H. Raffy, K. Kadowaki, D. Hinks, L. Ozyuzer and J. C. Campuzano, Nature Physics {\bf 2}, 447 (2006).
\bibitem{arcs} W. S. Lee, I. M. Vishik, K. Tanaka, D. H. Lu, T. Sasagawa, N. Nagaosa, T. P. Deveraux, Z. Hussain, Z.-X. Shen, Nature {\bf 450}, 81 (2007).
\bibitem{pdj1} H.-B. Yang, J. D. Rameau, P. D. Johnson, T. Valla, A. M. Tsvelik, G. Gu, Nature {\bf 456}, 77 (2008).

\bibitem{oscillations} N. Doiron-Leyraud, C. Proust, D. LeBoeuf, J. Levallois, J.-B. Bonnemaison, R. Liang, D. A. Bonn, W. N. Hardy, and L. Taillefer, Nature (London) {\bf 447}, 565 (2007).
\bibitem{pockets1} E. A. Yelland, J. Singleton, C. H. Mielke, N. Harrison, F. F. Balakirev, B. Dabrowski, and J. R. Cooper, Phys. Rev. Lett. {\bf 100}, 047003 (2008).

\bibitem{chubmorr} A. V. Chubukov and D. K. Morr,  Phys. Rep. {\bf 288}, 355 (1997).
\bibitem{sedr} T. A. Sedrakyan and A. V. Chubukov, Phys. Rev. B{\bf 81}, 174536 (2010).

\bibitem{japan} T. Tomeno, T. Machi, K. Tai, N. Kashizuka, S. Kambe, A. Hayashi, Y. Ueda and H. Yasuoka, Phys. Rev. B{\bf 49}, 15327 (1994).
\bibitem{bobroff} J. Bobroff, H. Alloul, P. Mendels, V. Viallet, J.-F.Marucco and D. Colson, Phys. Rev. Lett. {\bf 78}, 3757 (1997).
\bibitem{itoh} Y. Itoh, T. Machi, S. Adachi, A. Fukuoka, K. Tanabe and H. Yasuoka, J. Phys. Soc. Jpn. {\bf 67}, 312 (1998).



\bibitem{demler} Y. Zhang, E. Demler and S. Sachdev, Phys. Rev. B{\bf 66}, 094501 (2002).
\bibitem{tranquada} J. M. Tranquada, H. Woo, T. G. Perring, {\it et. al.} J. Phys. Chem. Solids {\bf 67}, 511 (2006).

\bibitem{loshad} I. E. Dzyaloshinskii, Phys. Rev. B{\bf 68}, 085113 (2003).
\bibitem{chubdash} B. L. Altshuler, A. V. Chubukov, A. Dashevskii, A. M. Finkelstein, D. Morr,
EPL  {\bf 41}, 401 (1998).
\bibitem{EssTsv} F. H. L. Essler and A. M. Tsvelik, Phys. Rev B{\bf 65}, 115 117 (2002);  Phys. Rev B{\bf  71}, 195116 (2005).
\bibitem{kotliar1} S. S. Kancharla, B. Kyung, D. S\'en\'echal, M. Civelli, M. Capone, G. Kotliar and A.-M. Tremblay, Phys. Rev. B{\bf 77}, 184516 (2008).
\bibitem{kotliar2} M. Ferrero, P. S. Cornaglia, L. De Leo, O. Parcollet, G. Kotliar and A. Georges, Phys. Rev. B {\bf 80}, 064501 (2009).
\bibitem{Imada} S. Sakai, Y. Motome, M. Imada, arXiv:1004.2569.
\bibitem{krt} R. M. Konik, T. M. Rice and A. M. Tsvelik, Phys. Rev. Lett. {\bf 96}, 086407 (2006).
\bibitem{RiceJohn} K.-Y. Yang, H.-B. Yang, P. D. Johnson, T. M. Rice and F.-Ch. Zhang, EPL {\bf 86}, 37002 (2009).
\bibitem{carbotte} J. P. F. Leblanc, J. P. Carbotte and E. J. Nicol, arXiv: 0910.3577.
\bibitem{subir} Y. Qi and S. Sachdev, Phys. Rev. B{\bf 81}, 115129 (2010).
\bibitem{moon} E. G. Moon and S. Sachdev, Phys. Rev. B{\bf 80}, 035117 (2009).
\bibitem{sandvik} A. W. Sandvik, Phys. Rev. Lett. {\bf 104}, 177201 (2010).
\bibitem{khodas} M. Khodas and A. M. Tsvelik, Phys. Rev. B{\bf 81}, 155102 (2010).
\end{thebibliography}
\end{document}